\documentclass[conference,compsoc]{IEEEtran}
\usepackage{makeidx}

\ifCLASSOPTIONcompsoc
  \usepackage[nocompress]{cite}
\else
  \usepackage{cite}
\fi
\ifCLASSINFOpdf
  \usepackage[pdftex]{graphicx}
\else
\fi
\usepackage{algorithm}
\usepackage{algpseudocode}

\usepackage{color}
\usepackage{comment}

\newcommand{\eat}[1]{}

\begin{document}
%
% paper title
% Titles are generally capitalized except for words such as a, an, and, as,
% at, but, by, for, in, nor, of, on, or, the, to and up, which are usually
% not capitalized unless they are the first or last word of the title.
% Linebreaks \\ can be used within to get better formatting as desired.
% Do not put math or special symbols in the title.
\title{Efficient Training of Convolutional Neural Nets on Large Distributed Systems}

% author names and affiliations
% use a multiple column layout for up to three different
% affiliations
\author{\IEEEauthorblockN{Sameer Kumar, Dheeraj Sreedhar}
\IEEEauthorblockA{IBM Research - India\\
Bangalore, KA, India, 560045\\
sameerk@us.ibm.com\\
dhsreedh@in.ibm.com}
\and
\IEEEauthorblockN{Vaibhav Saxena, Yogish Sabharwal\\
and Ashish Verma}
\IEEEauthorblockA{IBM Research - India\\
New Delhi, India, 110070\\
vaibhavsaxena@in.ibm.com\\
ysabharwal@in.ibm.com\\
vashish@in.ibm.com}}

% conference papers do not typically use \thanks and this command
% is locked out in conference mode. If really needed, such as for
% the acknowledgment of grants, issue a \IEEEoverridecommandlockouts
% after \documentclass

% for over three affiliations, or if they all won't fit within the width
% of the page (and note that there is less available width in this regard for
% compsoc conferences compared to traditional conferences), use this
% alternative format:
% 
%\author{\IEEEauthorblockN{Michael Shell\IEEEauthorrefmark{1},
%Homer Simpson\IEEEauthorrefmark{2},
%James Kirk\IEEEauthorrefmark{3}, 
%Montgomery Scott\IEEEauthorrefmark{3} and
%Eldon Tyrell\IEEEauthorrefmark{4}}
%\IEEEauthorblockA{\IEEEauthorrefmark{1}School of Electrical and Computer Engineering\\
%Georgia Institute of Technology,
%Atlanta, Georgia 30332--0250\\ Email: see http://www.michaelshell.org/contact.html}
%\IEEEauthorblockA{\IEEEauthorrefmark{2}Twentieth Century Fox, Springfield, USA\\
%Email: homer@thesimpsons.com}
%\IEEEauthorblockA{\IEEEauthorrefmark{3}Starfleet Academy, San Francisco, California 96678-2391\\
%Telephone: (800) 555--1212, Fax: (888) 555--1212}
%\IEEEauthorblockA{\IEEEauthorrefmark{4}Tyrell Inc., 123 Replicant Street, Los Angeles, California 90210--4321}}

% use for special paper notices
%\IEEEspecialpapernotice{(Invited Paper)}

% make the title area
\maketitle

% As a general rule, do not put math, special symbols or citations
% in the abstract
\begin{abstract}
Deep Neural Networks (DNNs) have achieved impressive accuracy in many application domains including image classification.
Training of DNNs is an extremely compute-intensive process and is solved using variants of the stochastic
gradient descent (SGD) algorithm.
A lot of recent research has focussed on improving the
performance of DNN training.
In this paper, we present optimization techniques to improve the performance of the data parallel synchronous SGD algorithm using the Torch framework: (i) we maintain  data in-memory to avoid file I/O overheads, (ii) we present a multi-color based MPI Allreduce algorithm to minimize communication overheads, and (iii) we propose optimizations to the Torch data parallel table framework that handles multi-threading.
We evaluate the performance of our optimizations on a Power 8 Minsky cluster with 32 nodes and 128 NVidia Pascal P100 GPUs.
With our optimizations, we are able to train 90 epochs of the 
      ResNet-50 model on the Imagenet-1k dataset using 256 GPUs in just 48 minutes.
      This significantly improves on the previously best known performance of training 
      90 epochs of the ResNet-50 model on the same dataset using 256 GPUs in 
      65 minutes. To the best of our knowledge, this is the best known training
      performance demonstrated for the Imagenet-1k dataset.

\end{abstract}

% no keywords

% For peer review papers, you can put extra information on the cover
% page as needed:
% \ifCLASSOPTIONpeerreview
% \begin{center} \bfseries EDICS Category: 3-BBND \end{center}
% \fi
%
% For peerreview papers, this IEEEtran command inserts a page break and
% creates the second title. It will be ignored for other modes.
\IEEEpeerreviewmaketitle

\section{Introduction}
%--> Background on deep learning
%--> Talk about parallelization efforts very briefly
%--> Discuss different frameworks
%--> Discuss challenges in parallelization (particularly I/O and synchronization aspects)
%--> We can discuss some of the prior work that is directly related (e.g. Facebook effort)
%--> Our contributions:
     % + Maintaining In memory distributed database and shuffling
     % + MPI Optimizations
     % + Other optimizations briefly
     % + Mention impactful results obtained including beating the facebook results.

Deep Neural Networks (DNNs) have achieved impressive accuracy in 
many application domains such as image classification and localization,
object detection, speech recognition and video classification\cite{alexnet_nips_12,dnn_speech,KarpathyCVPR14}.
In particular, the image classification challenge has resulted in the 
development of several deep neural networks such as AlexNet~\cite{alexnet_nips_12},
GoogleNet~\cite{googlenet_14}, VGG~\cite{vgg_19}, Resnet~\cite{residual_learning_15} and network in network
(NiN)~\cite{nin_LinCY13}. The imagenet dataset is a large scale dataset for image 
ontology that is frequently used in the research community. 
The Imagenet-1k dataset contains 1.2 million images and 1,000 categories and the
Imagenet-22k dataset contains 7 million images and 22,000 categories.

The objective of DNNs is to learn the weight vector that minimizes 
some measure of difference between the actual output and the predicted output.
Thus, a DNN essentially solves a non-convex optimization problem,
the most commonly used technique being the mini-batch Stochastic Gradient Descent (SGD).
In this approach, the input samples are split into small batches (mini-batches).
Each batch is then processed to calculate the gradients and update the model weights.
A pass through all the input images is typically referred to as an epoch. 
The algorithm executes multiple epochs (typically tens to hundreds)
iteratively until the desired accuracy is achieved.

Training of DNNs is an extremely compute-intensive process involving convolutions 
and matrix multiplications. Accelerators such as Graphic Processing Units (GPUs)
are quite efficient in handling such compute-intensive operations and are thus
widely used to accelerate the training process~\cite{dnn_speech,dan_arxiv2010}. Even with a GPU
it can take several days to train with certain models and datasets.
This has motivated researchers to explore efficient parallel and distributed 
SGD techniques in order to bring down the training time~\cite{downpour_nips_12,ibm_zhang,google_iclr_16}. 
The most common approach to parallelizing SGD is the synchronous data-parallel SGD,
wherein multiple workers orchestrate to process a mini-batch collectively.
Each worker maintains a copy of the latest model. 
One mini-batch is processed in every iteration by partitioning it amongst the parallel workers.
Each worker computes the gradients for its partition and then
the gradients across all the workers are accumulated to update the model weights. 
At the end of each iteration, every worker acquires the latest model with the updated weights.

The performance of synchronous data-parallel SGD algorithm depends on several factors.
One of the factors is the secondary storage I/O performance of the system. 
As the images are stored on secondary storage and a random set of images constituting 
the mini-batch have to be fetched in every iteration, there is significant file I/O overhead
due to random disk accesses.
One way to handle this issue is to augment the system with flash storage or other high performance
storage solutions; these are typically costly.
Another important factor that affects the training performance is the communication 
algorithm for accumulating the weights and updating the models on the workers. 
Note that as we scale to larger nodes, we would like to increase the mini-batch size 
as well so that the compute to communication ratio remains high and the 
compute resources are effectively utilized. However, from an accuracy perspective, 
there is a limit to the size of the mini-batch; increasing the mini-batch size beyond
a certain limit results in a drop in the accuracy. Therefore, on very large nodes we 
have to work with smaller batch size per worker. This leads to a lower compute to
communication ratio resulting in the communication algorithm becoming an important
factor in determining the training performance. 
The Message Passing Interface (MPI) programming model is used to
parallelize various parallel computing applications. It provides a rich
set of point-to-point and collective operations. In this framework, 
the gradients are accumulated using the MPI Allreduce collective. 
Many different algorithms have been proposed for optimizing
this collective in research literature~\cite{Thakur2005,ipdps16_coll, kumar_2013}.

Many frameworks have been developed to enable training and inferencing of deep learning models.
The most notable ones are Caffe~\cite{deep_scale_16}, Tensorflow from Google~\cite{tensor_flow_16}, Torch~\cite{torch-imagenet}, 
MXNet~\cite{mxnet_2015}, Chainer~\cite{chainer_learningsys2015} and Theano~\cite{theano_14}. Each framework has its own advantages.
Many of these frameworks support distributed training as well.

In this paper we present techniques to optimize the performance of the 
parallel synchronous SGD.
We have used the Torch framework since it is known to give a good balance between a high level scripting language and a low level 
compute efficient language. While the overall control can
be programmed in the scripting language LUA, all the heavy-lifting tasks can be implemented in C and interfaced to LUA through an
efficient foreign function interface (FFI) in Torch.
%{\color{red} Give some justification as to why?}.
We evaluate the performance improvements obtained using our optimizations on a POWER8 minsky
cluster interconnected using the Mellanox Infiniband~\cite{p8_redbook,connectx2_2010} network.
The main contributions of this paper are as follows:
\begin{itemize}
\item We present an in-memory data distribution strategy to overcome the file I/O
	  bottleneck. This optimization is particularly useful on systems exhibiting low
      file I/O performance. The strategy involves loading the entire data into the
      (distributed) memory of the system. An important aspect of this strategy is
      periodic shuffle operations that ensure randomness in the selection of the 
      mini-batches. In our experimental results, we demonstrate that this optimization
      results in 33\% improvement in the performance of GooglenetBN and 25\%
      improvement in performance of Resnet-50 models using the Imagenet-1k dataset.
\item We present an optimized multi-color algorithm for the MPI Allreduce collective
      that is used for accumulating the gradients from the workers.
      We show that this algorithm takes 50-60\% lesser time  in comparison to the MPI
      Allreduce implementation of the OpenMPI library.
\item We present optimizations to the Torch data parallel table implementation for
      application multi-threading. In our experimental results, we demonstrate that
      these optimizations lead to a performance improvement of 15-18\% for the
      GooglenetBN and Resnet-50 models.
\item Combining all these optimizations we are able to train 90 epochs of the 
      ResNet-50 model on the Imagenet-1k dataset using 256 GPUs in just 48 minutes.
      This significantly improves on the previously best known performance of training 
      90 epochs of the ResNet-50 model on the same dataset using 256 GPUs in 
      65 minutes. To the best of our knowledge, this is the best known training
      performance demonstrated for the Imagenet-1k dataset.
\end{itemize}

The rest of this paper is organized as follows.
In Section~2, we discuss work related to parallel SGD approaches for deep learning. In Section~3, we briefly present the data parallel distributed SGD framework; our optimizations are developed on this framework.
In Section~4, we present our algorithmic optimizations,
namely in-memory data strategy, MPI optimizations
and improvements to the Torch data parallel table infrastructure.
Next, in Section~5, we present results from our experimental evaluation of the proposed optimizations.
We finally present our conclusions in Section~6.

% eat

\section{Related Work}
% A more detailed discussion of prior work. Async, etc. should be covered here.
%
% Speedup and top-1 results on NiN, Googlenet, VGG
%
There has been significant amount of work to speed-up the training of deep neural networks on large distributed compute infrastructure. Specifically, it corresponds to training Stochastic Gradient Descent (SGD) algorithm in a data parallel way on a large number of nodes. There are multiple challenges involved towards this goal, such as, large communication overhead during synchronizing  the gradients across different nodes, using a large batch size which often results in lower accuracy, file I/O time, etc. 

The Message Passing Interface (MPI)~\cite{mpi-22,mpi-31}, a programming model to parallelize various computing applications, is often used for efficient communication of gradients across nodes. It provides a rich set of point-to-point and collective operations. Acceleration of distributed SGD with MPI calls has been presented before~\cite{elastic_nips_15,deep_scale_16}. An approach for distributed SGD with MPI is to designate one of the MPI processes as a parameter server that collects gradients from the peer processes and then sends updated gradients back to the peers~\cite{elastic_nips_15}. Other approaches take advantage of reduction trees via the peer-to-peer collective calls in MPI~\cite{deep_scale_16}. More details on the MPI usage and proposed improvements are discussed later in Sections~\ref{sec_parsgd} and \ref{multicolor}.

As the number of nodes increase, the effective batch size for SGD algorithm increases linearly which hampers the final accuracy which can be achieved. In~\cite{intel}, authors investigated the cause of poor performance of large batch sizes as compared to small batch sizes. They concluded that large batch size schemes results in sharp minimizers and so their generalization performance is not as good as the broad minimizers provided by small batch sizes. They experimentally validated this hypothesis by conducting several experiments with large and small batch sizes and report accuracy numbers along with the sharpness of the minimizers.
In~\cite{facebook}, authors proposed a new learning rate method in order to handle large batch-sizes (upto 8192) without affecting the accuracy. They also pipelined the computation and communication of gradient of different layers of the model to other nodes to minimize the impact of communication overhead. They were able to complete the training of Resnet50 model on Imagenet1k dataset in almost 1 hour, which was the best reported training time on 256 GPUs thus far. 

Another way to alleviate the problem of large batch size is to use asynchronous SGD. In the synchronous SGD case, the gradients from all the workers are aggregated across all the workers before the model parameters are updated. On the other hand, in the asynchronous case, model parameters are updated with a subset of workers without waiting for all the workers to finish. The main challenges to address is staleness of the gradients from workers and it's impact on convergence of the model parameters. A number of different approaches have been proposed in the literature to address these challenges~\cite{downpour_nips_12,ibm_zhang,elastic_nips_15,nips_2015,google_iclr_16}. However, synchronous SGD still seems to outperform various asynchronous approaches on large parallel systems and hence continue to be used for large model/dataset training of deep neural networks. 
 
\section{Data Parallel Distributed SGD}
% Present the parallel SGD algorithm here with pseudocode
\label{sec_parsgd}
In this section we briefly describe the data-parallel 
distributed SGD algorithm; our optimizations are based on this algorithm, outlined in Algorithm~\ref{algo_parsgd}.
We use the MPI+threads based programming model to parallelize synchronous SGD. We set up one MPI process per learner (compute node) where each learner further comprises of multiple GPU. Each GPU is driven by a separate thread within the MPI process. We used Torch7~\cite{torch-ppc64} and modified the Torch-MPI package~\cite{mpiT} to implement this data parallel distributed SGD algorithm.

The model weights ($W$) are initialized on each GPU  with identical random weights. 
In each training iteration, an effective batch of size $B$ is selected
for training in a distributed manner.
Let $\ell$ be the number of learners (nodes).
Each learner randomly samples a set of $B_{node}=B/N$ images
from the training data using a different random number seed.
These $B_{node}$ images are then equally divided across $m$ GPUs within a node such that each GPU works on a sample of $B_{GPU}=B_{node}/m$ images. 
Each GPU then performs a forward and backward computation to compute the gradients. The gradients from all the GPUs within a node are accumulated by performing a local intra-node summation. In the next step, a global inter-node summation and update of gradients across all the learners is performed using MPI Allreduce collective routine. These updated gradients on a node are broadcast to all the GPUs within the learner. Finally, each GPU performs SGD to update its copy of the model weights using the received updated gradients. 

\begin{algorithm}[ht]
\caption{Data-parallel Distributed SGD}
\label{algo_parsgd}
\begin{algorithmic}
{
%\dontprintsemicolon
%\BlankLine
\small

\State $X$: Training data  
\State $W$: Model Weights  
\State $T$:  Number of training iterations  
\State $N$: Number of learners (compute nodes)
\State $m$: Number of GPUs on each node \\

%\BlankLine
\State Initialize W with identical random values on all GPUs

\For {t = 1 to T}  
\State      Randomly sample $B_{node}$ images from $X$ on each learner (node i
\State      Divide $B_{node}$ images equally amongst $m$ GPUs on a node  
\State      Compute gradient $\Delta W^{(t)}_{ij}$  on GPU j of node i
\State      Local intra-node summation on node i:  
\State     $$\Delta W^{(t)}_i = \sum_ j^m  \Delta W^{(t)}_{ij}$$  
\State      Global inter-node summation using MPI Allreduce:
\State    $$\Delta W^{(t)} = \sum_ i^N \Delta W^{(t)}_i$$  
\State      Broadcast $\Delta W^{(t)}$ to all the GPUs within a node  
\State      Perform SGD on each GPU to compute $W^{(t+1)}$ 
\EndFor

%\BlankLine
}
\end{algorithmic}
\end{algorithm}
\normalsize
 
\section{Parallelization and Optimizations}
In this section, we describe our optimizations for the data parallel distributed SGD algorithm based on the in-memory distribution of the training data and an optimized algorithm for MPI Allreduce.
\subsection{In Memory Data Distribution}
% How we maintain the distributed database
% Shuffling method 
% with pseudocode
\label{dimd}
We analyzed the performance of our Torch based data parallel distributed SGD implementation.
We observed that with 4x NVIDIA PASCAL GPUs on a single node, a critical scaling bottleneck
was insufficient I/O throughput from the file system. The Torch
donkeys (work-stealing worker threads) were unable to load the next samples of the
mini-batch before the GPUs finished executing the SGD computation on
the previous mini-batch samples. This severely limited the image processing
throughput on 4x P100 GPUs on a single node.
%in the POWER8 SMP node.

In order to overcome this bottleneck,
we developed a novel Distributed In-Memory Data (DIMD) strategy for the storing training and validation images. First we resized
the images such that shorter dimension is of size 256 and the larger 
size is chosen preserving the aspect ration. Such resizing is 
commonly used in deep learning experiments and known to have
negligible impact on recognition accuracy. The resized images are compressed 
and concatenated into two large files for the training and
validation data sets, respectively.  
In order to allow for efficient random access to any image, 
we also maintain an index file
which contains the start location of each image along with its
label id for both these data sets. 
The cumulative size of both these sets is about 74GB for Imagenet-1k \cite{ilsvrc_12} and 300 GB for Imagenet-22k.

If there is sufficient memory on each node, then the entire dataset
can be stored in its memory, otherwise the data needs to be
partitioned and only a subset of the data is loaded on each node. If the data is partitioned then each learner on a node does not have the view of the entire data set.
So we also designed a novel in-memory shuffle operation 
that randomly shuffles the images across the partitions.
This can be invoked after every fixed number of training steps 
to ensure that the batch selection is fairly random.
The API's developed for the DIMD can be summarized as follows.

%\begin{enumerate}

%\item 
{\emph \bf i) Partitioned Load.} With this API each learner on a node loads
a sub-set of the dataset into memory. The size of the
sub-set is based on the available memory at each
node. We can divide the learners into groups such that each group
of learners collectively own the entire dataset.
In one extreme, if there is enough memory available on the learners,
each learner can hold the entire data set; thus each learner would define a group.
On the other extreme, if there is limited memory on the learners,
each learner would hold $1/{\ell}$ of the data (where ${\ell}$ is
the number of learners) and all the learners would collectively form 
a group.

%\item 
{\emph  \bf ii) Random in-memory batch load.} With this API, 
each learner can fetch a random batch (a set of images along 
with its label) from memory.

%\item 
{\emph \bf iii) Shuffle data across learners (nodes).} This API is used 
to shuffle the partitions across the nodes. This is 
implemented using the MPI AllToAllV collective call. For
group based partitions, the shuffle could be restricted
to that particular group. This could be efficiently implemented
using the communicator group in MPI.

%\end{enumerate}

\begin{figure}
\begin{center}\includegraphics[width=3.5in]{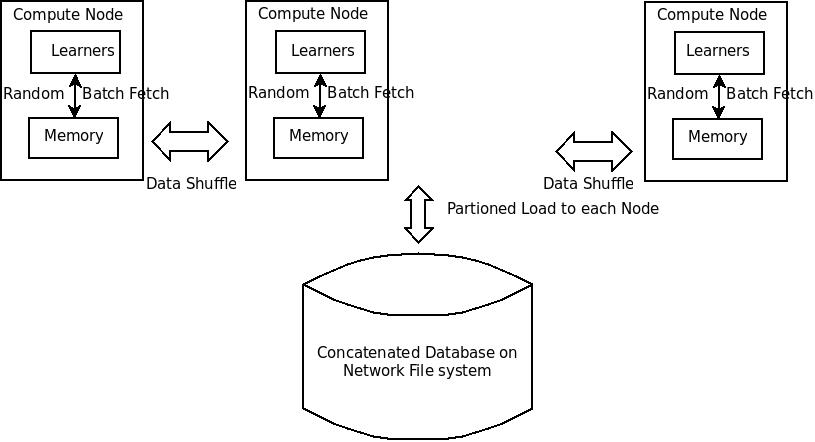}
\end{center}
\caption{Distributed in-memory database and the 3 API's for
managing it}
\label{fig:dimd}
\end{figure}

The processing of an epoch with the DIMD strategy is illustrated in Figure \ref{fig:dimd}. The in-memory shuffle algorithm is 
described in Algorithm \ref{algo_suff1}. When the entire
data set can fit in the main memory of each node, then the 
data partitioning can be achieved by each node having a partitioned set
of indices. The shuffle in this case just generation of random 
permutation of the indices. During the SGD execution, an in-memory JPEG decompresser
is also used to decompress images to generate image tensor objects that are used in the CUDA enabled Torch compute kernels.

\begin{algorithm}[ht]
\caption{Distributed in memory shuffle}
\label{algo_suff1}
\begin{algorithmic}
{
\small
\State $X$: Training data as a tensor of size $N \times 3 \times W \times H$
\State $N,W,H$ : no. of images, width and height
\State $r$: Rank of the current node
\State $S$: Size of the group in which current node belongs
\State $X$ = loadPartition(r,S) // loads from network file server
\State $m$ : no. of partitions, Partition $X$ into $m$ segments. // \emph{this is to
overcome the deficiency of MPI to handle more than 32 bit offsets}
\For {t = 1 to m}
\State $X_t$ = the $t$th sub-tensor from $X$ 
\State Partition $X_t$ into $S$ segments
\State Calculate the lengths and offsets of each segment
\State Exchange lenghts and offsets with every node
\State Perform AllToAllv on $X_t$ 
\EndFor
\State let $X^{'}$ be the collected AlltoAll output of all segments and let $N^{'}$ be the
number of images at current node after exchange
\State Shuffle $X^{'}$ within the node :
\State $perm$ = random permutation($1,..,N^{'}$)
\For {t = 1 to $N^{'}$}
\State $X^{''}(i)$  = $X^{'}(perm(t))$
\EndFor
%\State $X^{''}$ is the shuffled data set
%\BlankLine
}
\end{algorithmic}
\end{algorithm}
\normalsize

\subsection{Multi Color MPI Collective Algorithms}
%\subsubsection{Multi Color Collective Algorithms}
\label{multicolor}
We customized the MPI library to get optimal performance on our platform. In this section we describe our optimizations to the MPI Allreduce collective based on the multicolor collective algorithm.

\begin{figure}
\begin{center}
\includegraphics[width=3in]{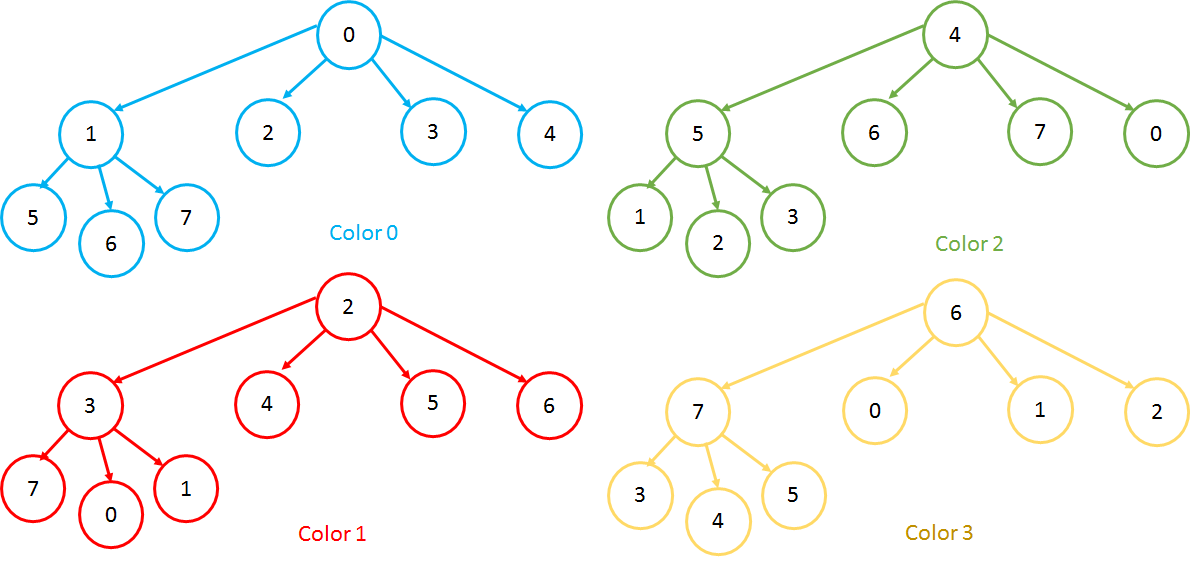}
\end{center}
\caption{4-color 4-ary trees on 8 Nodes. Note non leaf nodes are distinct across colors}
\label{fig:mctree}
\end{figure}

Multi-color collective algorithms~\cite{ipdps16_coll, kumar_2013} are tree-based algorithms
that execute the MPI collective operation along several different
paths of the network. In the k-color Allreduce algorithm, the
application payload is split into k chunks. Each of the k chunks is
summed along a different spanning tree that contains all the
nodes. The k-color Allreduce algorithm executes k pipelined reductions
to k different roots and then broadcasts the result from those roots
to the peer nodes. For each color, the nodes are arranged in a k-ary
BFS tree. While the tree paths can share network links, the non-leaf
nodes in the tree are disjoint among the
colors. Figure~\ref{fig:mctree} shows four spanning trees in a
4-color Allreduce performed on 8 nodes. The root and the non-leaf nodes perform
summing operations for the color, while the leaf nodes only send chunks
to the parent nodes. In Figure~\ref{fig:mctree}, chunk-0 is summed on
the tree color-0 rooted at node 0 with node 1 as the only non-leaf
node. Similarly, chunk-1 is summed on the tree color-1 rooted at node
2 with node 3 as the only non-leaf node. The network packets for each
color are transferred concurrently without any synchronization with other
colors, resulting in high throughput. 

On a fat-tree network, the multiple colors enable each color to use
different links from the root to the non-leaf nodes enabling
concurrent progress on the network. {If mapped to consecutive nodes on
the fat-tree network each non-leaf nodes of color will also push the
reductions and broadcasts to near neighbors resulting in minimization of
network contention}. However, we have also observed good link
utilization with nodes arbitrarily mapped on to the fat-tree.

Our implementation of the multi-color algorithm makes calls to
Infiniband verbs to initiate remote direct memory (RDMA) read
operations to enable nodes to pull data from the downstream nodes on
the tree. Direct calls to Infiniband verbs enable low latency and higher level
of pipelining on the reduction trees. We also use the PowerPC altivec
instruction set to sum network buffers with the local contributions
from the GPUs.

%\subsubsection{MPI Acceleration of SGD}
%\input{sgd}

\subsection{Data-Parallel Table Optimizations}
% Dynamic batch sizes
% What else ??

\begin{figure*}
\begin{center}
\hbox{\hspace{0.5em} \includegraphics[width=6.9in]{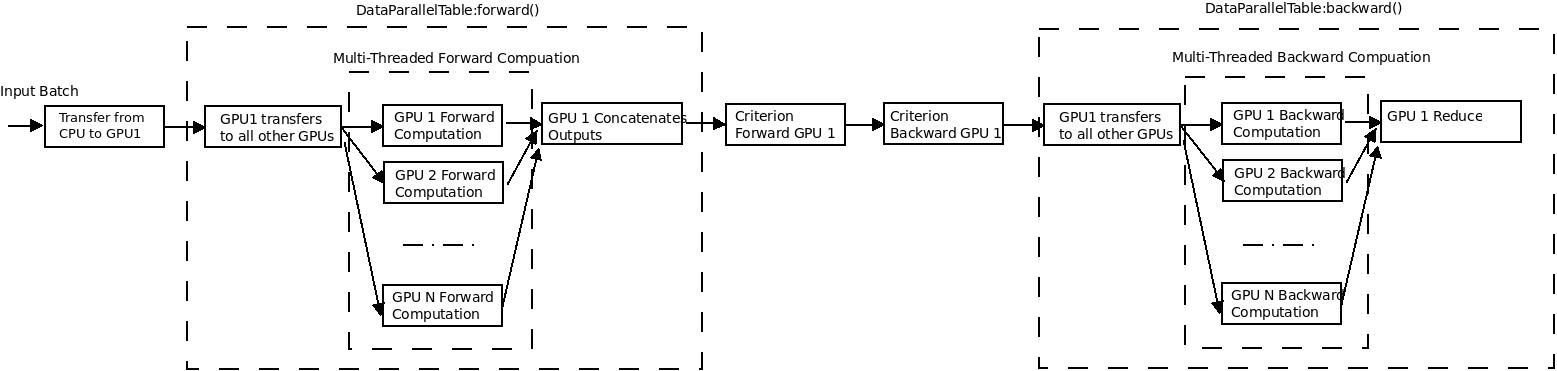}}
\end{center}
\caption{Data-Parallel-Table Implementation in the current Torch
framework which parallelizes the backward-forward computation on multiple GPUs attached to the same node}
\label{fig:dp_orig}
\end{figure*}

The Data-Parallel module \cite{torch-dp} in the Torch framework is responsible for the parallelization of work amongst the different GPU's attached to the
same SMP node. This module schedules the various cuDNN kernels involved
in the forward and backward computation of every training step.
The scheduling of multiple GPUs is done using the threading framework within Torch. Threads are created only once during the
initialization and jobs are submitted to the threading system by specifying a job function and an ending callback function. The job is subsequently executed on the first free thread. The ending callback function is executed in the main thread, when the job finishes - it is fully
serialized. While this technique is very elegant and easy to use,
the ending callback serialization does have overheads and it is best
to minimize its use. The current data-parallel table implementation
in Torch is shown in Figure \ref{fig:dp_orig}.

The main advantage of this implementation is that the same forward()
implementation can be used for training as well as inferencing. However, we have identified the following drawbacks with this design:

\begin{enumerate}
\item the entire input batch is first moved to the first GPU (GPU1) and then
partitioned on to the other GPUs. This results in more data movement as well
as more memory usage for the first GPU.
\item criterion evaluation is not parallelized.
\item the torch thread implementation results in more serialization.
\end{enumerate}

\begin{figure*}
\begin{center}
\hbox{\hspace{0.5em} \includegraphics[width=6.9in]{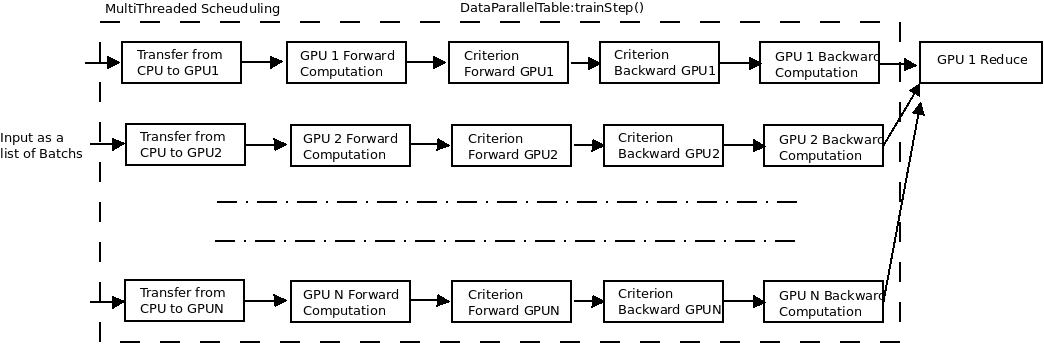}}
\end{center}
\caption{Optimized Data-Parallel-Table Implementation}
\label{fig:dp_opt}
\end{figure*}

To overcome, these shortcomings we re-designed the data-parallel table 
as shown in Figure \ref{fig:dp_opt}. The input batch is partitioned at the starting itself and is separately transfered to each GPU. Criterion evaluation is part of the data-parallel table and gets executed on every GPU. The number of serialization steps have been reduced.

\section{Experimental Evaluation}
%
% Performance results on NIN 
%
%
%\section{Experimental Evaluation}

We ran our distributed Torch application on a POWER8 Minsky cluster
interconnected by Infiniband two ConnectX-5 adapters each capable of a
raw bi-directional throughput of 100Gbps. Each POWER8 Minsky node has
20 processor cores, 256GB of host memory and four NVIDIA Pascal P100
GPUs. We ran the batch-normalized Googlenet \cite{googlenet_2} (GoogleNetBN) and ResNet-50 network models
available in the open-source Torch packages~\cite{torch-imagenet,
torch-resnet}. For all the experiments we used scale and aspect ratio data augmentation as in \cite{torch-resnet}. The input image is a $224 \times 224$ pixel random crop from a scaled image or its horizontal flip.
The input image is normalized by the per-color mean and
standard deviation.

We followed the warm start learning-rate schedule in \cite{facebook}.
The starting learning rate was fixed at 0.1. This is linearly ramped 
to $0.1 \frac{kn}{256}$, where $k$ is the batch size per GPU and $n$ is the the total number of workers (number of nodes times number of GPUs per node). We use a 90 epoch training regime with the learning rate dropped by a factor of 10 after every 30 epochs. We use a batch size of 64 per GPU for all the experiments unless otherwise specified.

We first independently study the performance improvements with each of our optimizations.
We first evaluate the MPI optimizations followed by our novel DIMD approach and finally the data-parallel table optimizations in the Torch framework. We then study the total improvement we get by combining all these optimizations and compare this with the state of the art
training performance.

\subsection{Effect of MPI Optimizations}
\begin{figure}
\begin{center}
\includegraphics[width=2.6in]{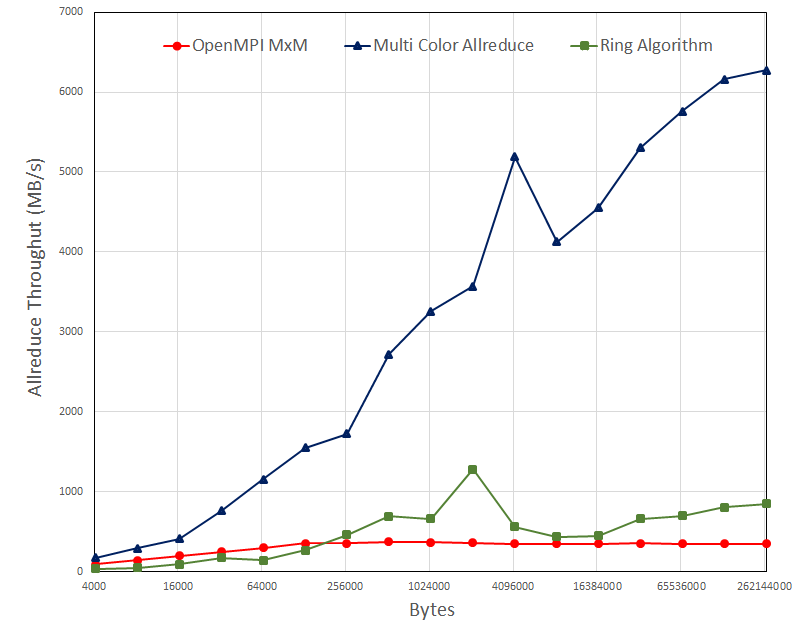}
\end{center}
\caption{MPI\_Allreduce throughout of various collective algorithms from CPU buffers}
\label{fig:aredthr}
\end{figure}

First we evaluate our MPI algorithm described in Section \ref{multicolor}, and compare it with ring-based and default OpenMPI algorithms. Figure ~\ref{fig:aredthr} shows the throughput of a
4-color MPI\_Allreduce on 16 POWER8 Minsky nodes with 64 GPUs. These
nodes were connected by 2 Mellanox ConnectX-5 adapters each with a
bi-directional raw link throughput of 100 Gbps. We also implemented a
pipelined ring algorithm where packets are reduced to a singe root
node along the ring then broadcast from the root to all peers in the
opposite direction.  The throughput achieved with the ring algorithm
is also presented in~Figure~\ref{fig:aredthr}. The performance of
the default OpenMPI algorithm is also presented. Note the multi-color
algorithm outperforms both the default OpenMPI algorithm and the ring
algorithm.
%%
%% compare algorithms
%%

\begin{figure}
\begin{center}
\includegraphics[width=2.6in]{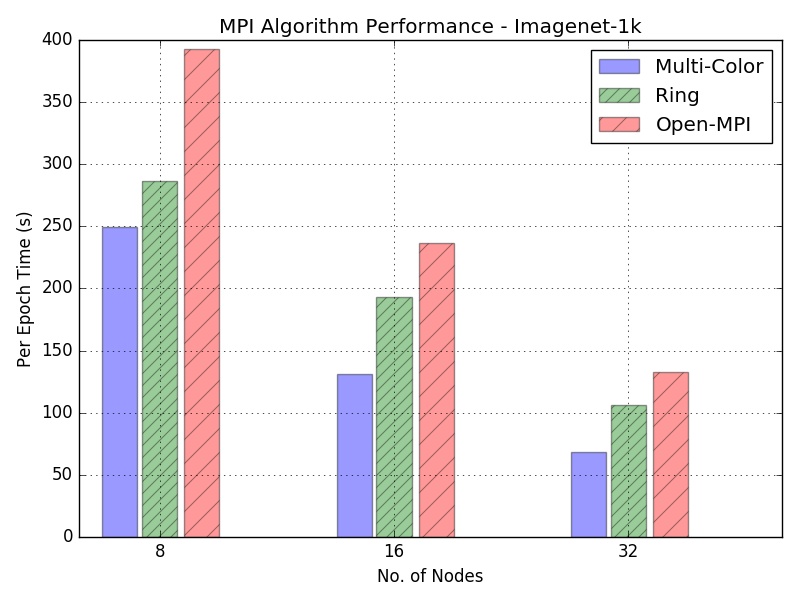}
\end{center}
\caption{Epoch time in seconds at various node counts with different MPI Allreduce schemes in our distributed Torch application}
\label{res:epoch_time}
\end{figure}

In the next experiment, we evaluate the overall training performance when using different MPI algorithms. We use GoogleNetBN with a reduction payload of 93MB for this experiment. 
The time to process one epoch is plotted for 8,16 and 32 learners in Figure \ref{res:epoch_time}. We observe that the multi-color algorithm takes 50-60\% lesser time than the default OpenMPI. It can also be observed that all the three algorithms scale with the number of learners but the multi-color algorithm gives the best scaling efficiency of 90.5\%. 

\subsection{Effect of DIMD Strategy}
\begin{figure}
\begin{center}
\hbox{\hspace{0.5em} \includegraphics[width=2.6in]{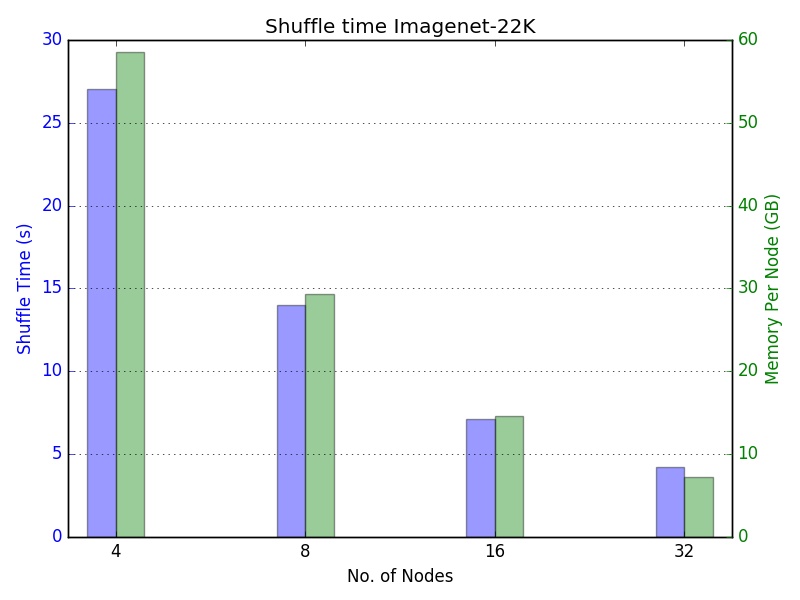}}
\end{center}
\caption{Time required for shuffle along with 
average memory utilization for each node for Imagenet-22k dataset}
\label{res:shuffle_22}
\end{figure}

\begin{figure}
\begin{center}
\hbox{\hspace{0.5em} \includegraphics[width=2.6in]{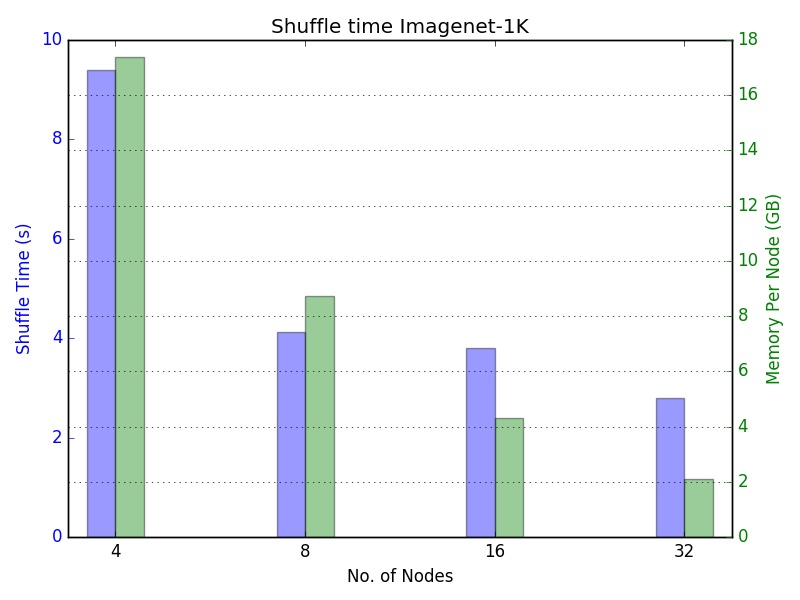}}
\end{center}
\caption{Time required for the shuffle along with 
average memory utilization for each node for Imagenet-1k dataset}
\label{res:shuffle_1}
\end{figure}

Next, we evaluate the performance advantages of our novel DIMD I/O technique. First we study the efficiency of our data set shuffle operation.
We evaluate the performance using two datasets, Imagenet-1k and
Imagenet-22k. The training dataset of Imagenet-1k has 1000 classes
and has 1.2 million images in total. Imagenet-22k has 22,000 classes
and has 7 million images. Using the concatenation technique 
described in Section \ref{dimd}, the training data set along with
the map indices of Imagenet-1k form a single file of size 70 GB and for
Imagenet-22k they form a single file of size 220 GB. First, we consider the
situation where the dataset is equally partitioned among all
the workers. 
The results are shown in
Figures \ref{res:shuffle_22} and \ref{res:shuffle_1}.
The shuffle-time along with the average memory
utilization is plotted for 8, 16 and 32 learners. 
We observe that the time required to shuffle the data amongst the learners decrease with increasing
number of learners. Note that as the number of learners
is doubled, the data held by each learner
is halved but now the data needs to be exchanged amongst double the number of learners. 
For Imagenet-22k the time to shuffle the entire
data among 32 learners is just 4.2 seconds.

\begin{figure}
\begin{center}
\hbox{\hspace{0.5em} \includegraphics[width=2.6in]{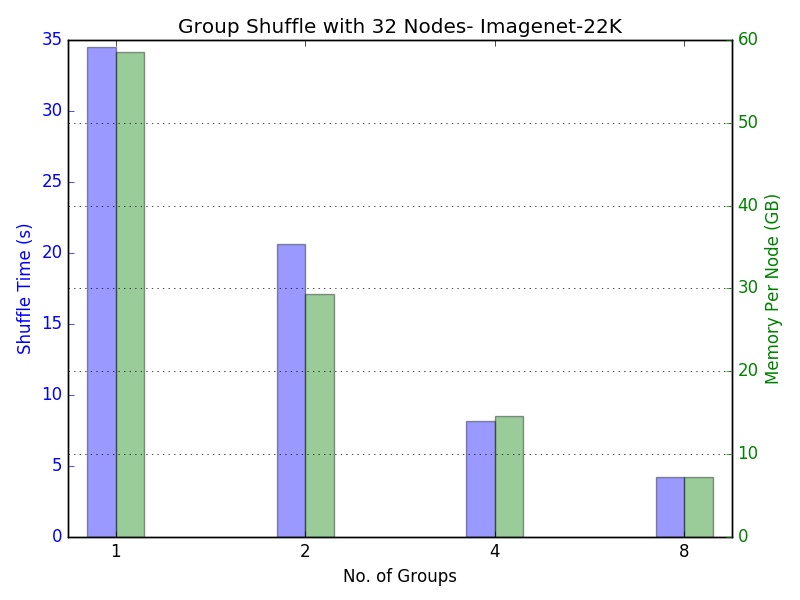}}
\end{center}
\caption{Time required for the group based shuffle along with 
average memory utilization for each node for Imagenet-22k dataset on 32 nodes.}
\label{res:shuffle_group}
\end{figure}

We now evaluate the performance of group-based shuffle when the data set is partitioned
among a subset (group) of learners, i.e.,  a subset of learners collectively own the entire dataset. We consider the Imagenet-22k dataset with 32 learners split
into 1, 4, 8 and 16 groups. The time to shuffle is plotted in 
Figure \ref{res:shuffle_group}. It can be observed that there is
not much improvement with the group based shuffle (compared to single
group). 
This is explained by the fact that all the connections are symmetrical in the cluster that we did our experiments on. Group
based shuffles are expected to give performance gains over non-group
based shuffle when locality can be exploited to create groups 
that are better connected amongst themselves.

\begin{figure}
\begin{center}
\hbox{\hspace{0.5em} \includegraphics[width=2.6in]{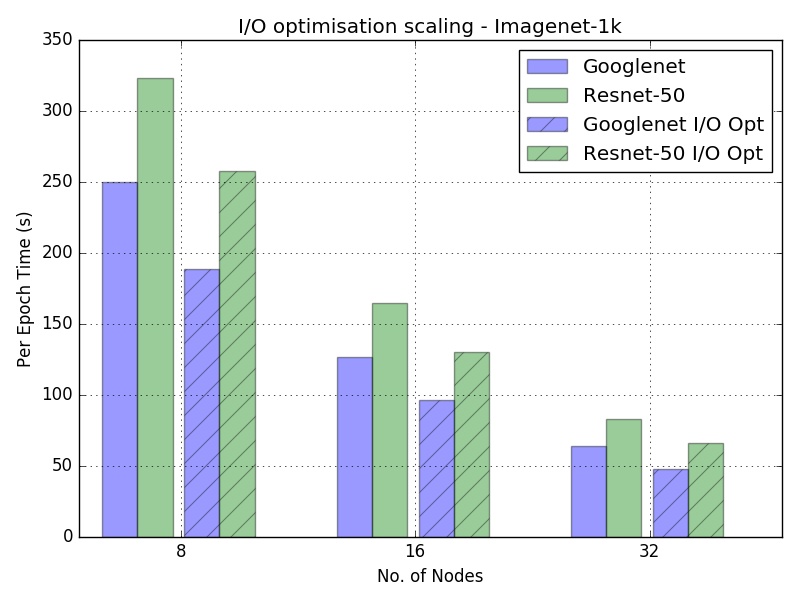}}
\end{center}
\caption{Epoch time in seconds at various node counts with and
without DIMD optimization for Imagenet-1k in our distributed Torch application.}
\label{res:io_1k}
\end{figure}

\begin{figure}
\begin{center}
\hbox{\hspace{0.5em} \includegraphics[width=2.6in]{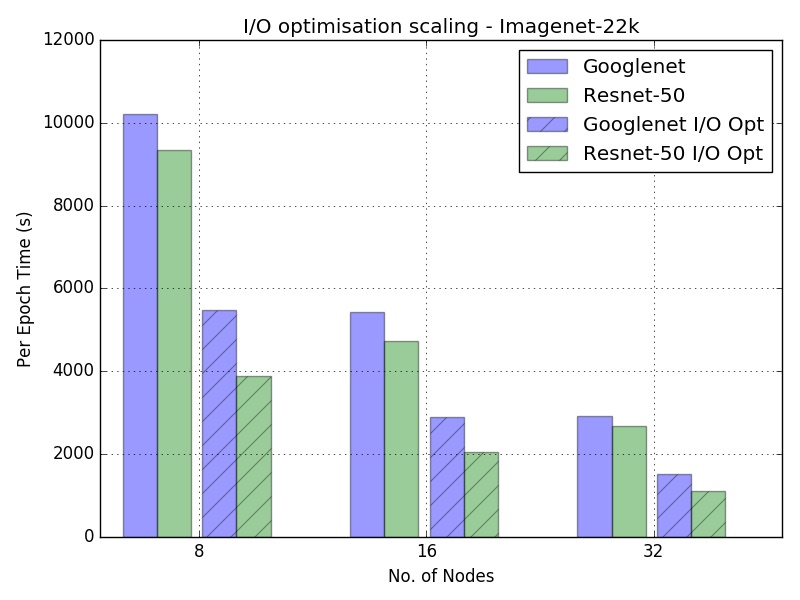}}
\end{center}
\caption{Epoch time in seconds at various node counts with and
without DIMD optimization for Imagenet-22k in our distributed Torch application.}
\label{res:io_22k}
\end{figure}

Next, we study the performance gains of DIMD on the overall training throughput. 
We use the optimal multi-color ring algorithm
for reduction in these experiments. The time to process one epoch for 8,16 and 32 learners with and without DIMD optimizations are plotted in Figures \ref{res:io_1k} and \ref{res:io_22k}.We present results for
both GooglenetBN and Resnet-50. For Imagenet-1k dataset, the proposed DIMD optimizations 
improve the per-epoch time for GooglenetBN by 33\% 
and Resnet-50 by 25\%. 
%The improvement in scaling is marginal.

\subsection{Effect of Data Parallel Table Optimizations}
\begin{figure}
\begin{center}
\hbox{\hspace{0.5em} \includegraphics[width=2.6in]{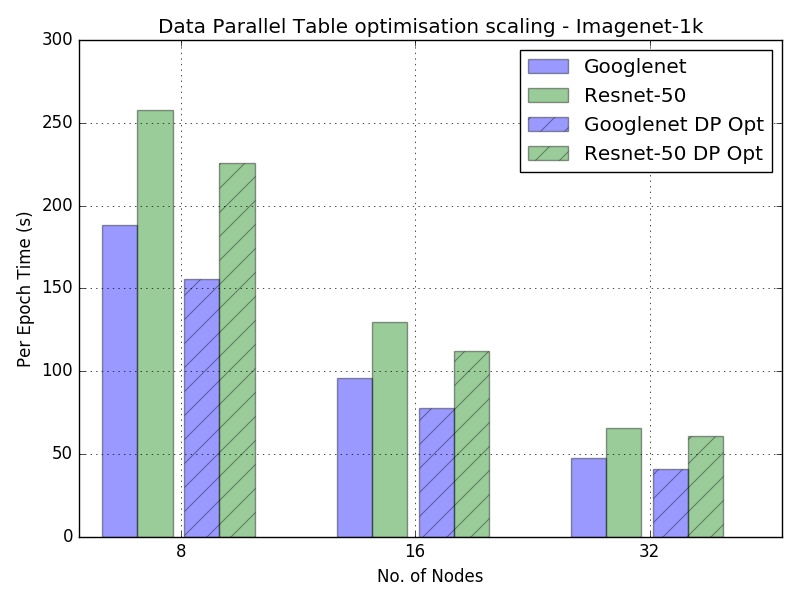}}
\end{center}
\caption{Epoch time in seconds at various node counts with and
without data-parallel optimizations in our distributed Torch application.}
\label{res:dp_1k}
\end{figure}

We next evaluate the performance improvements with our Data parallel table optimizations. 
For these experiments, we use the multi-color ring algorithm for reduction along with the DIMD optimizations. The time to process one epoch for 8, 16 and 32 learners with and without the data parallel table optimizations are plotted in Figure \ref{res:dp_1k}. We present results for both GooglenetBN and Resnet-50. 
We see that the proposed optimizations 
improve the per-epoch time by 15\% for GooglenetBN and by 18\% for Resnet-50. The improvement in scaling is marginal.

\subsection{Accuracy Evaluation}
\begin{figure}
\begin{center}
\includegraphics[width=3.2in]{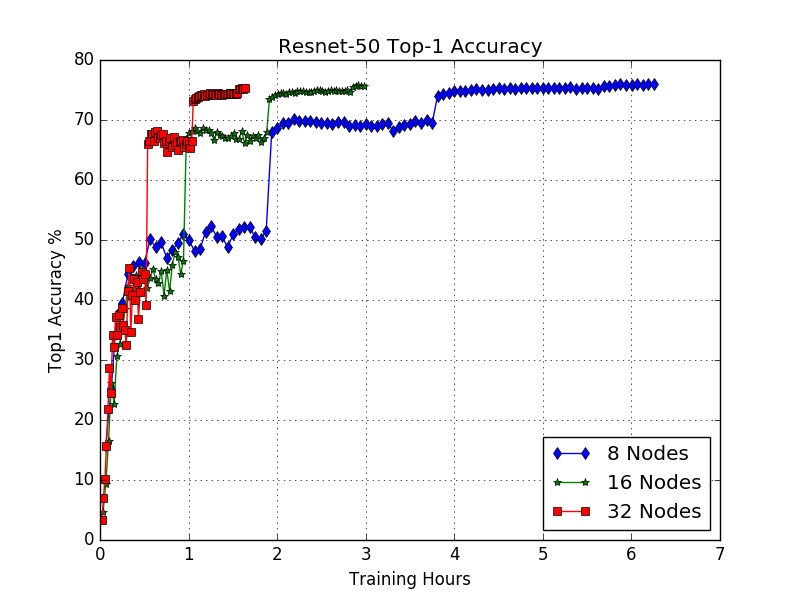}
\end{center}
\caption{Validation top-1 accuracy achieved over time in hours on different node counts for Resnet-50}
\label{res:acc_res}
\end{figure}

\begin{figure}
\begin{center}
\includegraphics[width=3.2in]{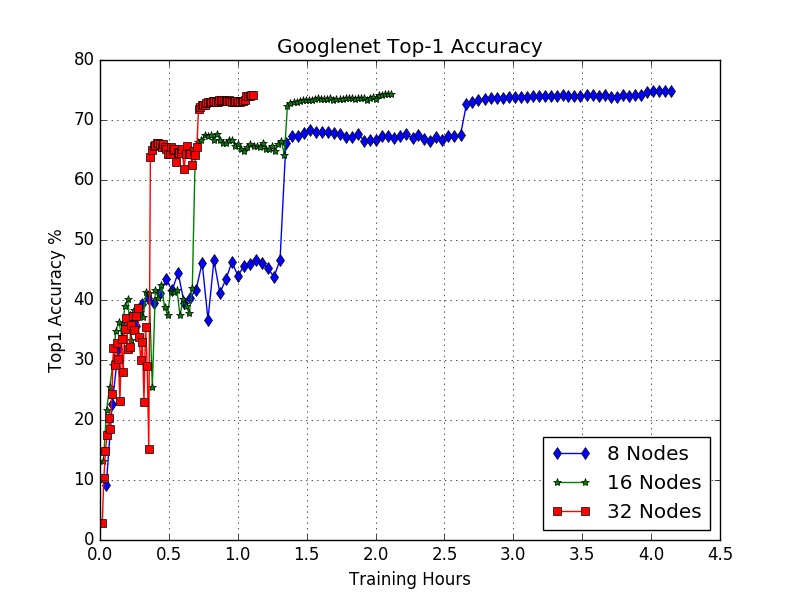}
\end{center}
\caption{Validation top-1 accuracy achieved over time in hours on different node counts for GooglenetBN}
\label{res:acc_googlenet}
\end{figure}

\begin{figure}
\begin{center}
\includegraphics[width=3.2in]{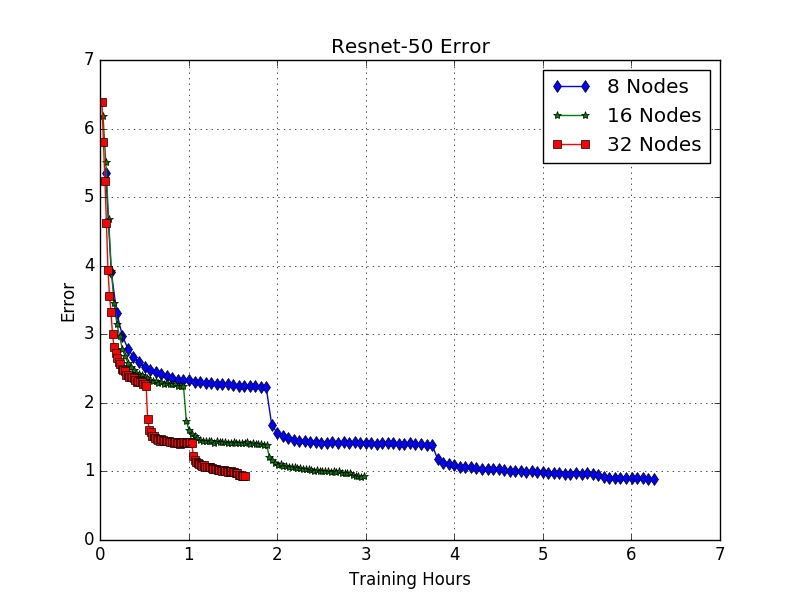}
\end{center}
\caption{Error as function of training time in hours on different node counts for Resnet-50}
\label{res:err_res}
\end{figure}

\begin{figure}
\begin{center}
\includegraphics[width=3.2in]{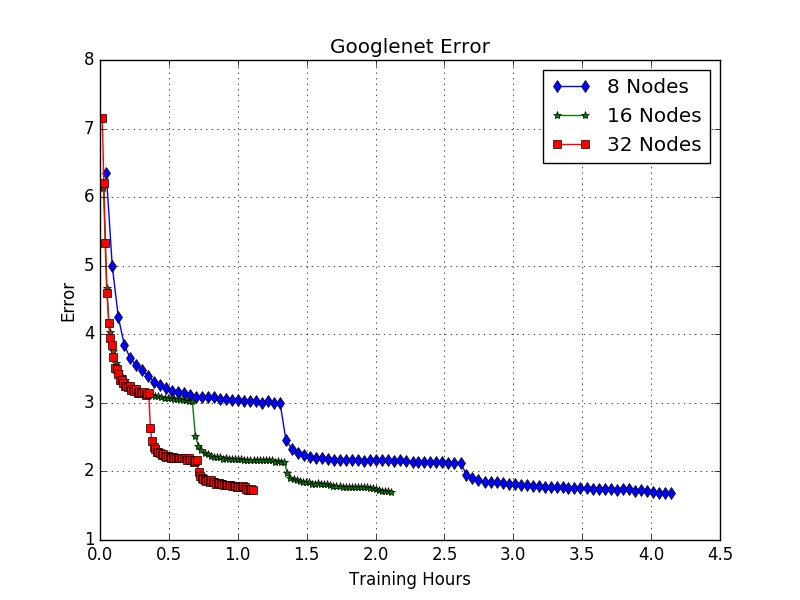}
\end{center}
\caption{Error as function of training time in hours on different node counts for Googlenet}
\label{res:err_googlenet}
\end{figure}

In this section, we finally evaluate the accuracy results of our distributed classifier. 
We evaluate the Top-1 validation accuracy for 8, 16 and 32 node runs; this is the percentage of the number of times the topmost predicted output is correct for Imagenet-1k. The accuracy results are plotted as a function of training time. Resnet-50 results are presented in Figure~\ref{res:acc_res} and GooglenetBN results in Figure~\ref{res:acc_googlenet}. In Figures~\ref{res:err_res}
 and \ref{res:err_googlenet} we present the objective function error as a function of training time for Resnet-50 and GooglenetBN respectively.
We note that none of the optimizations we presented have
any impact on the final accuracy of the classier. 
These results are presented to ensure correctness and completeness.

\subsection{Total Improvement and Comparison with state-of-the-art}
In Table \ref{tab:summary}, we summarize the total improvement
obtained with our contributions. We take the open source code 
of \cite{torch-imagenet} as the base. It should be be noted that
 there is no distributed Torch code available in open source.
 We created a distributed version with publicly available openMPI.
 This is the base version for our comparisons. We show improvements
 in the range of 58-72\% for GooglenetBN and in the range 110-130\% for
 Resnet-50. The peak accuracy obtained is also shown for reference.

\begin{table*}
\begin{center}
\begin{tabular}{c|c|c|c|c|c}
Model & Nodes & Time per epoch (s)   &  Time per epoch (s)  & Speedup   & Accuracy\\ 
&  & open source & fully Optimised &  &   \\  \hline
GooglenetBN & 8  & 249  & 155  & 60\%  & 74.86\% \\ \hline 
GooglenetBN & 16 & 131 & 76 &  72\% &   74.36\% \\ \hline 
GooglenetBN & 32  & 65 & 41 &  58\% &   74.19\%  \\  \hline 
Resnet-50 & 8  & 498 & 224  &  120\% &  75.99\%  \\ \hline 
Resnet-50 & 16  & 251 & 109  &  130\% &  75.78\% \\ \hline 
Resnet-50 & 32  & 128 & 58 &  110\% &  75.56\% \\ \hline 
\vspace{0.2cm}
\end{tabular}
\caption{Summary of total performance improvement along with the peak accuracy obtained for the classifier. The base version is \cite{torch-imagenet} along with the publicly available OpenMPI}
\label{tab:summary}
\end{center}
\end{table*}

In Table \ref{tab:comp} we compare our results with 
some of the state-of-the-art results,
\cite{facebook} and \cite{intel2}, on the same data sets. 
In \cite{facebook}, the authors use the same hardware as the one used in this paper, Nvidia P100 GPUs. With 256 P100 GPUs, they
were able to run 90 epochs of Resnet-50 in 65 minutes. The peak accuracy obtained was 76.2\%. With 256 P100 GPUs we could completed 90 epochs in just 48 minutes and reach a peak accuracy 75.4\%. We used a batch size of 32 per GPU for this experiment. We note that the difference between the two implementations are only in hyper-parameters; from a computation perspective, both are identical.
In \cite{intel2}, the authors present a distributed implementation using Intels Knights Landing (KNL)
processor. With 512 KNL processors they were able to complete 90
epochs in 60 mins and reach a peak accuracy of 74.7\%.

\begin{table*}
\begin{center}
\begin{tabular}{c|c|c|c|c|c}
Description & Hardware & Epochs & Batchsize & Accuracy & Time \\ \hline
Priya {\emph et al } \cite{facebook}  &  256 P100  & 90 & 8k & 76.2 \% 
& 65 mins \\ \hline
You  {\emph et al } \cite{intel2} & 512 KNL & 90 & 32k & 74.7 \% & 60 mins \\ \hline
Our work & 256 P100 & 90 & 8k & 75.4 \% & 48 mins\\ \hline
\vspace{0.2cm}
\end{tabular}
\caption{Comparisons with  \cite{facebook} and \cite{intel2}}
\label{tab:comp}
\end{center}
\end{table*}

\section{Conclusions and Future work}
We described our optimization techniques to improve the performance of data-parallel synchronous SGD algorithm used in distributed DNN training. Our techniques included an in-memory data distribution strategy to overcome the file I/O bottleneck, an optimized multi-color algorithm for the MPI Allreduce collective, and optimization to the Data Parallel Table module in Torch. 

Some of our optimization techniques, such as in-memory data distribution, are generic enough to be used in other Deep Learning frameworks for improving the training time on systems that exhibit slow file I/O performance. 

In future, we would like to explore the use and impact of our optimizations for the case of asynchronous SGD. For example, in-memory data distribution technique should also improve the data loading performance in the asynchronous case and yield an overall gain in the training performance. However, being an asynchronous setting, different methods have to be designed to handle data shuffle, etc. 
 
%%\section{POWER8 Systems Architecture}
%%\input{smpi_p8.tex}

%\section{Summary}
%\input{summary}

% use section* for acknowledgment
%\ifCLASSOPTIONcompsoc
  % The Computer Society usually uses the plural form
%  \section*{Acknowledgments}
%\else
  % regular IEEE prefers the singular form
%  \section*{Acknowledgment}
%\fi

%The authors would like to thank...

% trigger a \newpage just before the given reference
% number - used to balance the columns on the last page
% adjust value as needed - may need to be readjusted if
% the document is modified later
%\IEEEtriggeratref{8}
% The "triggered" command can be changed if desired:
%\IEEEtriggercmd{\enlargethispage{-5in}}

% references section

% can use a bibliography generated by BibTeX as a .bbl file
% BibTeX documentation can be easily obtained at:
% http://mirror.ctan.org/biblio/bibtex/contrib/doc/
% The IEEEtran BibTeX style support page is at:
% http://www.michaelshell.org/tex/ieeetran/bibtex/
%\bibliographystyle{IEEEtran}
% argument is your BibTeX string definitions and bibliography database(s)
%\bibliography{IEEEabrv,../bib/paper}
%
% <OR> manually copy in the resultant .bbl file
% set second argument of \begin to the number of references
% (used to reserve space for the reference number labels box)

\bibliographystyle{IEEEtran}
\bibliography{IEEEabrv,references}

\begin{comment}

\end{comment}

% that's all folks
\end{document}